\documentclass{revtex4}
\usepackage{amsmath}
\usepackage{amssymb}
\usepackage{graphicx}
\usepackage{bm}
\usepackage{dcolumn}

\newcommand{\bpm}{\begin{pmatrix}}
\newcommand{\epm}{\end{pmatrix}}

\renewcommand{\Re}{\mathop{{\rm Re}}\nolimits}

\begin{document}

\title{Weakly incoherent magnetotransport in layered metals}
\author{P. D. Grigoriev}
\email{grigorev@itp.ac.ru}
\affiliation{L. D. Landau Institute for Theoretical Physics, Chernogolovka, Russia}
\date{\today }

\begin{abstract}
We investigate electronic conductivity in layered metals in
magnetic field in the weakly incoherent limit, when the interlayer
transfer integral is smaller than the Landau level broadening due
to the impurity potential, but the interlayer electron tunnelling
conserves the intralayer momentum. It is shown that the impurity
potential has much stronger effect in this regime, than in the
quasi-2D metals in the coherent limit. The weakly incoherent
regime has several new qualitative features, not found in the
previous theoretical approaches. The background interlayer
magnetoresistance in this regime monotonically grows with
increasing of magnetic field perpendicular to the conducting
layers. The effective electron mean free time is considerably
shorter than in the coherent regime and decreases with magnetic
field. This enhances the role of higher harmonics in the angular
magnetoresistance oscillations and increases the Dingle
temperature, which damps the magnetic quantum oscillations.
\end{abstract}

\maketitle

\section{Introduction}

The crossover between coherent and incoherent electron transport
in the layered metals attracts great attention, both theoretical
\cite{MosesMcKenzie1999,Abrikosov1999,Lundin2003,OsadaIncoherent,Ho,Gvozd2007,Maslov,Incoh2009}
and
experimental,\cite{Incoh2009,Zuo1999,Wosnitza2002,CrossoverNAture2002,MarkPRL2006,Ardavan2006,Zverev}
for its influence on the properties of high-temperature cuprate
superconductors, organic metals, heterostructures, and many other
layered materials. This crossover can be driven by temperature
$T$, volume impurity concentration $n_{i}$, external magnetic
field $\mathbf{B}=\left( B_{x},B_{y},B_{z}\right) $. In magnetic
field this crossover in conductivity is very pronounced because it
qualitatively changes the magnetoresistance behavior.

The electronic conductivity in magnetic field is widely used to investigate
the electronic structure of various metals. In strongly anisotropic quasi-2D
layered metals, when the interlayer transfer integral $t_{z}$ is much
smaller than the Fermi energy $E_{F}$, the influence of magnetic field on
conductivity has many specific features. One has to separate several
different regimes of interlayer magnetotransport, depending on the ratios of
three energy parameters: the interlayer transfer integral $t_{z}$, the
inverse mean free time $\Gamma _{0}=\hbar /2\tau _{0}$ due to the impurity
scattering, and the Landau level (LL) separation $\hbar \omega _{c}$, where $%
\omega _{c}=eB_{z}/m^{\ast }c$ is the cyclotron frequency.

When the interlayer transfer integral is larger than the Landau
level separation, $t_{z}>\hbar \omega _{c}$, the 3D electronic
dispersion is well defined and given in the tight-binding
approximation by
\begin{equation}
\epsilon _{3D}\left( \mathbf{k}\right) \approx \epsilon \left(
k_{x},k_{y}\right) -2t_{z}\cos (k_{z}d),  \label{ES3D}
\end{equation}%
where $\epsilon \left( k_{x},k_{y}\right) $ is the in-plane
electron dispersion and $d$ is the interlayer spacing. Then the
classical magnetoresistance shows Yamaji
oscillations,\cite{Yam,Yagi1990} which are used to determine the
in-plane Fermi momentum. The magnetic quantum oscillations (MQO)
of interlayer conductivity in this case have beats of
amplitude,\cite{Shoenberg} and these beats are shifted with
respect to the beats of MQO of magnetization or of the other
thermodynamic quantities.\cite{PhSh,ShubCondMat,Shub} The slow
oscillations also appear in the interlayer conductivity, which can
be used to separate relaxation times from different scattering
mechanisms.\cite{Shub,SO}

When the interlayer transfer integral is smaller than the Landau
level separation, $t_{z}<\hbar \omega _{c}$, the beats of MQO
disappear. This limit happens in strong fields in very anisotropic
metals. If the interlayer transfer integral is still larger than
the LL broadening, $t_{z}>\Gamma $, the dispersion (\ref{ES3D})
survives, and the MQO can be described by the "coherent" theory in
Refs.
\cite{PhSh,ShubCondMat,Shub,SO,ChampelMineev,Gvozd2004,ChMineevComment2006}.
Note, that the LL broadening $\Gamma $ is larger than $\Gamma
_{0}$ in strongly anisotropic metals close to a stack of isolated
conducting layers [see Eq. (\ref{GB}) below].

In the very anisotropic dirty limit, when the interlayer transfer
integral is the smallest parameter, $t_{z}<\hbar \omega
_{c},\Gamma $, the traditional 3D approach fails to describe the
interlayer magnetoresistance. For example, in this limit, the
experimentally observed interlayer magnetoresistance grows with
increasing of the out-of-plane magnetic field strength $B$ not
only in the maxima, but also in the minima of MQO (see, e.g.,
Refs. \cite{Incoh2009,Zuo1999,Wosnitza2002,MarkPRL2006}). The
angular dependence of the background magnetoresistance also has
many unusual features in this regime.\cite{MarkPRL2006,Incoh2009}
This change of the magnetoresistance behavior as the magnetic
field strength or the impurity concentration increase was called
the "coherence-to-incoherence crossover". It has been observed in
various compounds and attracted the considerable theoretical
attention.\cite{MosesMcKenzie1999,Gvozd2007,Incoh2009} The term
"weakly incoherent" has been introduced\cite{MosesMcKenzie1999} to
separate this regime from the coherent 3D limit $t_{z}>\hbar /\tau
$, and from the completely incoherent regime, where the electron
tunnelling to the adjacent layers does not conserve the in-plane
electron momentum. The completely incoherent interlayer electron
tunnelling happens when it goes via resonance
impurities,\cite{Abrikosov1999,Maslov,Incoh2009} due to
interaction with phonons\cite{Lundin2003}\cite{Ho} and in some
other models.

The theory of weakly incoherent magnetoresistance in Ref. \cite%
{MosesMcKenzie1999} is based on the phenomenological Green function (see Eq.
(53) of Ref. \cite{MosesMcKenzie1999}), which is equivalent to%
\begin{equation}
G_{R}^{0}({\boldsymbol{r}}_{1},{\boldsymbol{r}}_{2},j,\varepsilon
)=\sum_{n,k_{y}}\frac{\Psi _{n,k_{y},j}^{0\ast }(x_{2},y_{2})\Psi
_{n,k_{y},j}^{0}(x_{1},y_{1})}{\varepsilon -\varepsilon _{n}-i\Gamma _{0}}.
\label{G0}
\end{equation}%
Here $j$ is the number of conducting layer, related to the $z$-coordinate as
$z=jd$. The Landau level number $n$ and the momentum component $k_{y}$ form
the complete set of quantum numbers of the 2D electrons in magnetic field
with free electron dispersion
\begin{equation}
\varepsilon _{n}=\hbar \omega _{c}\left( n+1/2\right) .  \label{En2D}
\end{equation}%
In magnetic field ${\boldsymbol{B}}=(B_{x},0,B_{z})$ (we may choose $B_{y}=0$
without loss of generality because the in-plane dispersion is uniform) the
electromagnetic potential in the Landau gauge is ${\boldsymbol{A}}%
=(zB_{y},xB_{z}-zB_{x},0)$. Then the 2D electron wave functions are%
\begin{equation}
\Psi _{n,k_{y},j}\left( x,y\right) =\Psi _{n}\left( x-l_{Hz}^{2}\left[
k_{y}+jd/l_{Hx}^{2}\right] \right) e^{ik_{y}y},  \label{Psi2D}
\end{equation}%
where%
\begin{equation}
\Psi _{n}\left( x\right) =\frac{\exp \left( -x^{2}/2l_{Hz}^{2}\right)
H_{n}\left( x/l_{Hz}\right) }{\left( \pi l_{Hz}^{2}\right) ^{1/4}2^{n/2}%
\sqrt{n!}},  \label{Psi2DH}
\end{equation}%
$H_{n}\left( x/l_{Hz}\right) $ is the Hermite polynomial and, for brevity,
we introduced the notation of magnetic length components $l_{Hx}=\sqrt{\hbar
c/eB_{x}}$ and $l_{Hz}=\sqrt{\hbar c/eB_{z}}$. In the Green function in Eq. (%
\ref{G0}) the impurity scattering produces only the imaginary part of the
self-energy $i\Gamma _{0}$, which is independent of the quantum numbers $%
\left\{ n,k_{y},j\right\} $ and of the magnetic field strength $B$. This
approximation is incorrect in the weakly incoherent regime, as will be shown
in Sec. II below.

The Green function in Eq. (\ref{G0}) is not suitable to study the
MQO, because the MQO of the electron density of state (DoS) at the
Fermi level lead to the similar oscillations of the electron
self-energy, which must be taken into account in the theory of
MQO.\cite{ShubCondMat,Shub,ChampelMineev,Gvozd2004,ChMineevComment2006}
For electrons with 3D dispersion, as in Eq. (\ref{ES3D}), in the
Born approximation and after averaging over the impurity
configurations, the imaginary part of the self-energy is
proportional to the density of states, i.e. it acquires the
oscillating energy dependence:
\begin{equation}
\Gamma =\Gamma \left( \varepsilon \right) =\Gamma _{0}\left[ 1+\rho \left(
\varepsilon ,B\right) /\rho _{0}\right] ,  \label{G}
\end{equation}%
where $\rho \left( \varepsilon \right) $ and $\rho _{0}$ are the electron
DoS with and without magnetic field. The electron Green function
\begin{equation}
G_{R}^{0}({\boldsymbol{r}}_{1},{\boldsymbol{r}}_{2},j,\varepsilon
)=\sum_{n,k_{y},k_{z}}\frac{\Psi _{n,k_{y},j}^{0\ast }(x_{2},y_{2})\Psi
_{n,k_{y},j}^{0}(x_{1},y_{1})e^{ik_{z}\left( z_{1}-z_{2}\right) }}{%
\varepsilon -\varepsilon _{2D}\left( n,k_{y}\right) +2t_{z}\cos
(k_{z}d)-i\Gamma \left( \varepsilon \right) }.  \label{GN2D}
\end{equation}%
with $\Gamma \left( \varepsilon \right) $ from Eq. (\ref{G}) has
been substituted to the Kubo formula in the calculation of MQO of
interlayer conductivity $\sigma _{zz}$ in quasi-2D metals in Refs.
\cite{ShubCondMat,Shub,ChampelMineev,Gvozd2004,ChMineevComment2006}
The completely incoherent hopping mechanism of the interlayer
magnetotransport, which does not conserve the in-plane electron
momentum during the interlayer hopping, has also been
suggested\cite{Gvozd2007} to explain the exponential growth in
interlayer magnetoresistance with decreasing temperature. However,
all these approaches cannot explain the monotonic increase of
magnetoresistance with increasing $B$ in the minima of
MQO, observed in $\beta ^{\prime \prime }$-(BEDT-TTF)$_{2}$SF$_{5}$CH$_{2}$CF%
$_{2}$SO$_{3}$.\cite{Wosnitza2002,Zuo1999}

Below we reexamine the approach based on Eqs.
(\ref{G0})-(\ref{GN2D}), in the weakly incoherent limit $\hbar
\omega _{c}>\Gamma _{0}>t_{z}$. We argue that Eq. (\ref{G}) does
not hold in this limit, and derive the different formula for the
Green function. Then we calculate the interlayer conductivity with
the new Green function and show that the new result considerably
differs from that in the "coherent" theory in Refs.
\cite{MosesMcKenzie1999,PhSh,ShubCondMat,Shub,SO,ChampelMineev,Gvozd2004,ChMineevComment2006}.
This explains several new qualitative features of MQO and of the
angular dependence of interlayer magnetoresistance observed in the
weakly incoherent limit.

\section{The model}

The electron Hamiltonian in layered compounds with small interlayer coupling
consists of the 3 terms%
\begin{equation}
\hat{H}=\hat{H}_{0}+\hat{H}_{t}+\hat{H}_{I}.  \label{H}
\end{equation}%
The first term $\hat{H}_{0}$ is the 2D free electron Hamiltonian summed over
all layers:%
\begin{equation*}
\hat{H}_{0}=\sum_{m,j}\varepsilon _{2D}\left( m\right) c_{m,j}^{+}c_{m,j},
\end{equation*}%
where $\left\{ m\right\} $\ is the set of quantum numbers of electrons in
magnetic field on a 2D conducting layer, $\varepsilon _{2D}\left( m\right) $%
\ is the corresponding free electron dispersion given by Eq. (\ref{En2D}),
and $c_{m}^{+}(c_{m})$ are the electron creation (annihilation) operators in
the state $\left\{ m\right\} $. The second term in Eq. (\ref{H}) gives the
coherent electron tunnelling between two adjacent layers:
\begin{equation}
\hat{H}_{t}=2t_{z}\int dxdy[\Psi _{j}^{\dagger }(x,y)\Psi _{j-1}(x,y)+\Psi
_{j-1}^{\dagger }(x,y)\Psi _{j}(x,y)],  \label{Ht}
\end{equation}%
where $\Psi _{j}(x,y)$ and $\Psi _{j}^{\dagger }(x,y)$\ are the creation
(annihilation) operators of an electron on the layer $j$ at the point $(x,y)$%
. We call this interlayer tunnelling Hamiltonian "coherent" because it
conserves the in-layer coordinate dependence of the electron wave function
(in other words, it conserves the in-plane electron momentum) after the
interlayer tunnelling. The last term\
\begin{equation}
\hat{H}_{I}=\sum_{i}V_{i}\left( r\right)  \label{Hi}
\end{equation}%
\ is the impurity potential. The impurities are taken to be point-like and
randomly distributed on the layers. The impurity distributions on any two
adjacent layers are uncorrelated. The potential $V_{i}\left( r\right) $ of
any impurity located at point $r_{i}$ is given by
\begin{equation}
V_{i}\left( r\right) =U\delta ^{3}\left( r-r_{i}\right) .  \label{Vi}
\end{equation}

In the 3D limit, when the interlayer transfer integral $t_{z}$ is
much larger than the electron level broadening $\Gamma $ due to
the impurity scattering, the impurity potential $\hat{H}_{I}$ can
be considered as the small perturbation for the electrons with
dispersion (\ref{ES3D}). In the Born approximation this gives
$\Gamma =\pi n_{i}U^{2}\rho \left( E_{F}\right) $ in agreement
with Eq. (\ref{G}), where $n_{i}$ is the volume impurity
concentration, and $\rho \left( E_{F}\right) $ is the DoS at the
Fermi level. This leads to the standard theory of magnetic quantum
oscillations in Q2D metals.\cite{Shoenberg}\cite{Shub} In the
opposite limit, $t_{z}\ll \Gamma ,\hbar \omega _{c}$, the
interlayer hopping $t_{z}$ must be considered as a perturbation
for the disordered uncoupled stack of 2D metallic layers, where
Eq. (\ref{G}) is no more
valid.\cite{Ando,Ando1,Baskin,Brezin,QHE,Imp,Burmi}

The 2D metallic electron system in magnetic field in the
point-like impurity potential has been extensively
studied.\cite{Ando,Ando1,Baskin,Brezin,QHE,Imp,Burmi} The
point-like impurity potential leads to the broadening of the
Landau levels, which is described by the density of states (DoS)
distribution function $D\left( E\right) $. Since each Landau level
without disorder is strongly degenerate, even weak impurity
potential lifts this degeneracy and leads to the considerable
broadening of the Landau levels. The electron Green functions
acquire a cut instead of the pole as in Eq. (\ref{G0}). In the
self-consistent one-site
approximation, the Green function is given by%
\begin{equation}
G({\boldsymbol{r}}_{1},{\boldsymbol{r}}_{2},\varepsilon )=\sum_{n,k_{y}}\Psi
_{n,k_{y}}^{0\ast }(r_{2})\Psi _{n,k_{y}}^{0}(r_{1})G\left( \varepsilon
,n\right) ,  \label{Gg}
\end{equation}
where%
\begin{equation}
G\left( E,n\right) =\frac{E+E_{g}\left( 1-c_{i}\right) \pm \sqrt{\left(
E-E_{1}\right) \left( E-E_{2}\right) }}{2EE_{g}},  \label{GAndo0}
\end{equation}%
and the DoS $D\left( E\right) =\left( -1/\pi \right) {\text{Im}}G_{R}\left(
E\right) $ on each LL is described by the dome-like function\cite{Ando}
\begin{equation}
D\left( E\right) =\frac{\sqrt{\left( E-E_{1}\right) \left( E_{2}-E\right) }}{%
2\pi \left\vert E\right\vert E_{g}},  \label{DoSAndo}
\end{equation}%
where the electron energy $E$ is counted from the last occupied LL, $%
E=\varepsilon -\varepsilon _{2D}\left( n,k_{y}\right) $, and%
\begin{equation*}
E_{g}=V_{0}/2\pi l_{Hz}^{2}.
\end{equation*}%
Here $V_{0}=U\left\vert \psi \left( z_{i}\right) \right\vert ^{2}\approx U/d$%
, is the 2D analogue of the strength $U$ of the point-like impurity
potential:
\begin{equation}
V_{i}\left( x,y\right) =V_{0}\delta \left( x-x_{i}\right) \delta \left(
y-y_{i}\right) ,  \label{Vi2D}
\end{equation}%
and $\psi \left( z\right) $ is the out-of-plane electron wave function. The
boundaries of the DoS dome in Eq. (\ref{DoSAndo}) are%
\begin{equation}
E_{1}=E_{g}\left( \sqrt{c_{i}}-1\right) ^{2},~E_{2}=E_{g}\left( \sqrt{c_{i}}%
+1\right) ^{2},  \label{E12}
\end{equation}%
where $c_{i}$ is the ratio of the 2D impurity concentration, $N_{i}=n_{i}d$,
to the 2D DoS on one LL, $N_{LL}=1/2\pi l_{Hz}^{2}$:
\begin{equation}
c_{i}=N_{i}/N_{LL}=2\pi l_{Hz}^{2}n_{i}d.  \label{ci}
\end{equation}%
The function in Eq. (\ref{DoSAndo}) is normalized to unity, $\int D\left(
E\right) dE=1$. $D\left( E\right) $ converges at the point $E=0$, because
this point lies outside the DoS dome $E_{1}<E<E_{2}$. The LL broadening
\begin{equation}
\Gamma _{B}\equiv \left( E_{2}-E_{1}\right) /2=2E_{g}\sqrt{c_{i}}\propto
\sqrt{B}.  \label{GE}
\end{equation}%
The ratio
\begin{equation}
\frac{\Gamma _{B}}{\Gamma _{0}}=\frac{2V_{0}\sqrt{n_{i}d/2\pi l_{Hz}^{2}}}{%
\pi n_{i}U^{2}\rho \left( E_{F}\right) }\approx \frac{2U\sqrt{n_{i}N_{LL}/d}%
}{\pi n_{i}U^{2}\rho \left( E_{F}\right) }=\frac{2\sqrt{n_{i}U^{2}\rho
\left( E_{F}\right) \hbar \omega _{c}}}{\pi n_{i}U^{2}\rho \left(
E_{F}\right) }=\sqrt{\frac{4\hbar \omega _{c}}{\pi \Gamma _{0}}}  \label{GEc}
\end{equation}%
grows as $\sqrt{B}$ in high magnetic field. Eqs. (\ref{GE}),(\ref{GEc}) give
the correct asymptotic for the LL broadening in strong magnetic field. In
weak magnetic field, when $\hbar \omega _{c}\ll \Gamma _{0}$, the mean
scattering time $\tau _{B}$ related to level broadening as $\Gamma
_{B}=\hbar /2\tau _{B}$ and entering the Drude formula, does not depend on
the value of magnetic field along the conductivity: $\tau _{B}=\tau _{0}$.%
\cite{Abrik} To get the correct asymptotic behavior for $\Gamma _{B}$ both
in strong magnetic field and at $B=0$, one can take the simple function
\begin{equation}
\Gamma _{B}\approx \Gamma _{0}\left[ \left( 4\hbar \omega _{c}/\pi \Gamma
_{0}\right) ^{2}+1\right] ^{1/4}.  \label{GB}
\end{equation}

More realistic models of the finite-range impurity potential, and more
accurate calculation of the DoS, including the many-site corrections, lead
only to the small tails of the DoS dome.\cite{Ando1}\cite{Brezin}\cite{Burmi}
The number of electron states in these tails is much less than the number of
states in the DoS dome and can be neglected. However, to include these tails
into account and to simplify the subsequent calculation, one can take the
Lorentzian DoS distribution with the same broadening:%
\begin{equation}
D\left( E\right) \approx \frac{\Gamma _{B}}{\pi \left( E^{2}+\Gamma
_{B}^{2}\right) }=-\frac{\text{Im}G_{R}\left( E\right) }{\pi }.
\label{LorentzDoS}
\end{equation}%
Combining Eqs. (\ref{Gg}), (\ref{Gn}) and (\ref{LorentzDoS}) we obtain
\begin{equation}
G({\boldsymbol{r}}_{1},{\boldsymbol{r}}_{2},\varepsilon )=\sum_{n,k_{y}}%
\frac{\Psi _{n,k_{y}}^{0\ast }(r_{2})\Psi _{n,k_{y}}^{0}(r_{1})}{\varepsilon
-\varepsilon _{n}-i\Gamma _{B}},  \label{GLor}
\end{equation}%
This Green function will be used in the next section to calculate the
interlayer conductivity.

\section{Calculation of conductivity.}

The interlayer conductivity $\sigma _{zz}$, associated with the Hamiltonian (%
\ref{Ht}), can be calculated using the Kubo formula and the formalism,
developed for the metal-insulator-metal junctions\cite{Mahan} The
conductivity is expressed via the electron on the adjacent layers:%
\begin{equation}
\sigma _{zz}=\frac{e^{2}t_{z}^{2}d}{\hbar L_{x}L_{y}}\left\langle \int d^{2}{%
\boldsymbol{r}}d^{2}{\boldsymbol{r}}^{\prime }\int \frac{d\varepsilon }{2\pi
}A({\boldsymbol{r}},{\boldsymbol{r}}^{\prime },j,\varepsilon )A({\boldsymbol{%
r}}^{\prime },{\boldsymbol{r}},j+1,\varepsilon )\left[ -n_{F}^{\prime
}(\varepsilon )\right] \right\rangle ,  \label{KuboA}
\end{equation}%
where the electron spectral functions
\begin{equation}
A({\boldsymbol{r}},{\boldsymbol{r}}^{\prime },j,\varepsilon )=i\left[ G_{A}({%
\boldsymbol{r}},{\boldsymbol{r}}^{\prime },j,\varepsilon )-G_{R}({%
\boldsymbol{r}},{\boldsymbol{r}}^{\prime },j,\varepsilon )\right] ,
\label{A}
\end{equation}%
and the advanced (retarded) Green's functions $G_{A(R)}({\boldsymbol{r}},{%
\boldsymbol{r}}^{\prime },j,\varepsilon )$ include interaction with
impurities. The product of two spectral functions in Eq. (\ref{KuboA})
rewrites as
\begin{gather}
\Pi \equiv A({\boldsymbol{r}},{\boldsymbol{r}}^{\prime },j,\varepsilon )A({%
\boldsymbol{r}}^{\prime },{\boldsymbol{r}},j+1,\varepsilon )=G_{A}({%
\boldsymbol{r}},{\boldsymbol{r}}^{\prime },j,\varepsilon )G_{R}({\boldsymbol{%
r}}^{\prime },{\boldsymbol{r}},j+1,\varepsilon )+  \label{PiA} \\
+G_{R}({\boldsymbol{r}},{\boldsymbol{r}}^{\prime },j,\varepsilon )G_{A}({%
\boldsymbol{r}}^{\prime },{\boldsymbol{r}},j+1,\varepsilon )-2\text{Re}G_{R}(%
{\boldsymbol{r}},{\boldsymbol{r}}^{\prime },j,\varepsilon )G_{R}({%
\boldsymbol{r}}^{\prime },{\boldsymbol{r}},j+1,\varepsilon ),  \notag
\end{gather}%
In addition to the terms with the product of Green functions $G_{A}G_{R}$,
the expression for conductivity also contains the term $-$Re$G_{R}G_{R}$,
which becomes important when MQO are considered.\cite{ChampelMineev,Shub}
Eq. (46) and subsequent formulas in Ref. \cite{MosesMcKenzie1999}, where
this term is omitted, can be applied only when MQO are disregarded. In
strong magnetic field especially in the layered metals, on contrary, the MQO
are very strong.

The angular brackets in Eq. (\ref{KuboA}) mean averaging over impurity
configurations. Since the impurity distributions on each layer is
uncorrelated with other layers, one can perform this averaging separately
for each spectral function independently, which gives:%
\begin{equation}
\sigma _{zz}=\frac{e^{2}t_{z}^{2}d}{\hbar L_{x}L_{y}}\int d^{2}{\boldsymbol{r%
}}d^{2}{\boldsymbol{r}}^{\prime }\int \frac{d\varepsilon }{2\pi }%
\left\langle A({\boldsymbol{r}},{\boldsymbol{r}}^{\prime },j,\varepsilon
)\right\rangle \left\langle A({\boldsymbol{r}}^{\prime },{\boldsymbol{r}}%
,j+1,\varepsilon )\right\rangle \left[ -n_{F}^{\prime }(\varepsilon )\right]
.  \label{KuboS}
\end{equation}%
The averaged Green (or spectral) functions are translational invariant: $%
\left\langle G_{R}({\boldsymbol{r}},{\boldsymbol{r}}^{\prime },j,\varepsilon
)\right\rangle =\left\langle G_{R}({\boldsymbol{r}}-{\boldsymbol{r}}^{\prime
},j,\varepsilon )\right\rangle $.

If the magnetic field is tilted by angle $\theta $ with respect to the
normal to the conducting planes, $B=\left( B_{x},0,B_{z}\right) =\left(
B\sin \theta ,0,B\cos \theta \right) $, the Green functions on two adjacent
layers acquire the phase shift [see Eq. (49) of Ref. \cite{MosesMcKenzie1999}%
]:%
\begin{equation}
G_{R}({\boldsymbol{r}},{\boldsymbol{r}}^{\prime },j+1,\varepsilon )=G_{R}({%
\boldsymbol{r}},{\boldsymbol{r}}^{\prime },j,\varepsilon )\exp \left\{ ie%
\left[ \Lambda \left( {\boldsymbol{r}}\right) -\Lambda \left( {\boldsymbol{r}%
}^{\prime }\right) \right] /\hbar \right\} ,  \label{GL}
\end{equation}%
where
\begin{equation*}
\Lambda \left( {\boldsymbol{r}}\right) =-yB_{x}d=-yBd\sin \theta .
\end{equation*}%
Substituting Eq. (\ref{GL}) into Eq. (\ref{PiA}) we obtain%
\begin{gather}
\Pi =2G_{A}({\boldsymbol{r}},{\boldsymbol{r}}^{\prime },j,\varepsilon )G_{R}(%
{\boldsymbol{r}}^{\prime },{\boldsymbol{r}},j,\varepsilon )\cos \left\{ e%
\left[ \Lambda \left( {\boldsymbol{r}}\right) -\Lambda \left( {\boldsymbol{r}%
}^{\prime }\right) \right] /\hbar \right\} -  \label{Pi} \\
-2\text{Re}\left[ G_{R}({\boldsymbol{r}},{\boldsymbol{r}}^{\prime
},j,\varepsilon )G_{R}({\boldsymbol{r}}^{\prime },{\boldsymbol{r}}%
,j,\varepsilon )\exp \left\{ -ie\left[ \Lambda \left( {\boldsymbol{r}}%
\right) -\Lambda \left( {\boldsymbol{r}}^{\prime }\right) \right] /\hbar
\right\} \right] .  \notag
\end{gather}%
and
\begin{eqnarray}
\sigma _{zz} &=&\frac{2e^{2}t_{z}^{2}d}{\hbar }\int d^{2}{\boldsymbol{r}}%
\int \frac{d\varepsilon }{2\pi }\left[ -n_{F}^{\prime }(\varepsilon )\right]
\times  \label{Kubo1} \\
&&\times \left\{ \left\vert \left\langle G_{R}({\boldsymbol{r}},\varepsilon
)\right\rangle \right\vert ^{2}\cos \left( \frac{eByd}{\hbar }\sin \theta
\right) \right.  \notag \\
&&-\left. \text{Re}\left[ \left\langle G_{R}({\boldsymbol{r}},\varepsilon
)\right\rangle ^{2}\exp \left( \frac{ieByd}{\hbar }\sin \theta \right) %
\right] \right\} .  \notag
\end{eqnarray}%
The term in the third line of Eq. (\ref{Kubo1}) is absent in Eq. (50) of
Ref. \cite{MosesMcKenzie1999}.

In the magnetic field perpendicular to the conducting layers%
\begin{equation}
\sigma _{zz}=\frac{2e^{2}t_{z}^{2}d}{\hbar }\int d^{2}{\boldsymbol{r}}\int
\frac{d\varepsilon }{2\pi }\left[ \left\vert \left\langle G_{R}({\boldsymbol{%
r}},\varepsilon )\right\rangle \right\vert ^{2}-\Re \left\langle G_{R}({%
\boldsymbol{r}},\varepsilon )\right\rangle ^{2}\right] \left[ -n_{F}^{\prime
}(\varepsilon )\right] .  \label{sperp}
\end{equation}%
The integration over ${\boldsymbol{r}}$ for the Green function of the form (%
\ref{Gg}) is very simple and gives%
\begin{equation}
\sigma _{zz}=\frac{2e^{2}t_{z}^{2}d\,N_{LL}}{\hbar }\int \frac{d\varepsilon
}{2\pi }\left[ -n_{F}^{\prime }(\varepsilon )\right] \sum_{n}\left[
\left\vert \left\langle G_{R}(\varepsilon ,n)\right\rangle \right\vert
^{2}-\Re \left\langle G_{R}(\varepsilon ,n)\right\rangle ^{2}\right] .
\label{sp1}
\end{equation}%
With the approximate Green function, given by Eq. (\ref{GLor}), Eq. (\ref%
{sp1}) becomes
\begin{equation}
\sigma _{zz}=\frac{2e^{2}t_{z}^{2}d\,N_{LL}}{\hbar }\int \frac{d\varepsilon
}{2\pi }\sum_{n}\frac{\left[ -n_{F}^{\prime }(\varepsilon )\right] 2\Gamma
_{B}^{2}}{\left[ \left( \varepsilon -\varepsilon _{n}\right) ^{2}+\Gamma
_{B}^{2}\right] ^{2}}.  \label{s2}
\end{equation}%
The sum and integral in Eq. (\ref{s2}) is calculated in a standard way,
transforming the sum over LL into the harmonic sum by applying the Poisson
summation formula:\cite{ZW}
\begin{equation}
\sum_{n=n_{0}}^{\infty }f(n)=\sum_{k=-\infty }^{\infty }\int_{a}^{\infty
}e^{2\pi ikn}f(n)\,dn
\end{equation}%
where $a\in (n_{0}-1;n_{0})$. Then, performing the integrations, we obtain
\begin{equation}
\sigma _{zz}=\sigma _{0}\left( B\right) \sum_{k=-\infty }^{\infty }\left(
-1\right) ^{k}\exp \left[ \frac{2\pi \left( ik\mu -\left\vert k\right\vert
\Gamma _{B}\,\right) }{\hbar \omega _{c}}\right] \frac{2k\pi ^{2}T/\hbar
\omega _{c}}{\sinh \left( 2k\pi ^{2}T/\hbar \omega _{c}\right) }\left[ 1+%
\frac{2\pi \left\vert k\right\vert \Gamma _{B}}{\hbar \omega _{c}}\right] .
\label{sLor}
\end{equation}%
where%
\begin{equation}
\sigma _{0}\left( B\right) =\frac{e^{2}t_{z}^{2}\nu _{F}d}{\hbar \Gamma _{B}}%
,  \label{s0}
\end{equation}%
$\nu _{F}=N_{LL}/\hbar \omega _{c}$ is the DoS at the Fermi level in the
absence of magnetic field. Eq. (\ref{sLor}) would coincide with Eqs.
(17)-(21) of Ref. \cite{ChampelMineev} if $\Gamma _{0}$ and $\Gamma
_{\varepsilon }$ in these equations are replaced by $\Gamma _{B}$. Note,
that the nonoscillating part of conductivity $\sigma _{0}\left( B\right) $
is now a function of magnetic field, because $\Gamma _{B}\propto \sqrt{B}$
in strong magnetic field. This observation contradicts the previous
theoretical results\cite{MosesMcKenzie1999}\cite{ChampelMineev}\cite%
{Gvozd2004}, also developed for the almost 2D case, because in these papers
the LL width $\Gamma _{B}$ was incorrectly taken to have no monotonic
dependence on magnetic field, $\Gamma _{B}=\Gamma _{0}+\tilde{\Gamma}$,
where $\tilde{\Gamma}$ rapidly oscillates around zero.

\begin{figure}[th]
\includegraphics[width=0.4\textwidth]{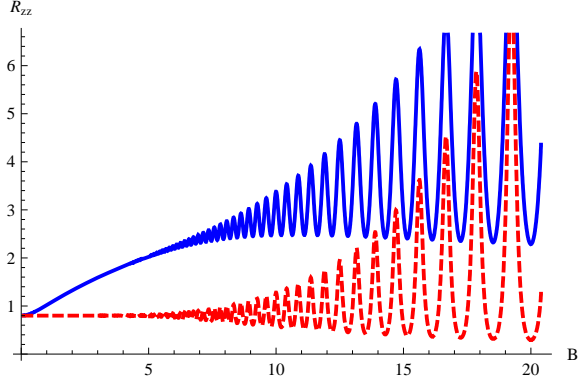}
\caption{{}The MQO of resistivity $R_{zz}\left( B\right) =1/\protect\sigma %
_{zz}$, calculated using Eq. (\protect\ref{sLor}) with $\Gamma _{B}$ given
by Eq. (\protect\ref{GB}) [solid blue line] and with $\Gamma _{B}=\Gamma
_{0} $ [dashed red line].}
\label{FigMQO}
\end{figure}

Let us now compare how strongly our result differs from the previous results
[see, e.g. Eqs. (17-21) of Ref. \cite{ChampelMineev}]. In Fig. \ref{FigMQO}
we plot the MQO of resistivity $R_{zz}\left( B\right) =1/\sigma _{zz}$,
calculated using Eq. (\ref{sLor}) with $\Gamma _{B}$ given by Eq. (\ref{GB})
[solid blue line] and with $\Gamma _{B}=\Gamma _{0}$ [dashed red line]. The
difference is evident: the interlayer magnetoresistance shows monotonic
growth and weaker oscillating amplitude with increasing magnetic field, than
in the old result.

In tilted magnetic field the calculation of Eq. (\ref{Kubo1}) performed in
Ref. \cite{MosesMcKenzie1999} can be applied with the new
magnetic-field-dependent value $\Gamma _{B}$ instead of $\Gamma _{0}$, which
gives [compare to Eq. (1) of Ref. \cite{MosesMcKenzie1999}]%
\begin{equation}
\sigma _{zz}=\sigma _{0}\left( B\right) \left\{ \left[ J_{0}\left( \kappa
\right) \right] ^{2}+2\sum_{\nu =1}^{\infty }\frac{\left[ J_{\nu }\left(
\kappa \right) \right] ^{2}}{1+\left( \nu \omega _{c}\tau _{B}\right) }%
\right\} ,  \label{sB}
\end{equation}%
where $\kappa \equiv k_{F}d\tan \theta $ and
\begin{equation}
\tau _{B}=\hbar /2\Gamma _{B}=\tau _{0}\left( \Gamma _{0}/\Gamma _{B}\right)
.  \label{tauB}
\end{equation}
There are two differences between this formula and Eq. (1) of Ref. \cite%
{MosesMcKenzie1999}. First, the higher harmonics in AMRO are weaker damped
in Eq. (\ref{sB}) because of smaller value of $\tau _{B}\propto 1/\sqrt{B}$.
Second, as we noted before, the background conductivity $\sigma _{0}\left(
B\right) $, given by Eq. (\ref{s0}), decreases as $1/\sqrt{B}$ in strong
field.

\begin{figure}[th]
\includegraphics[width=0.4\textwidth]{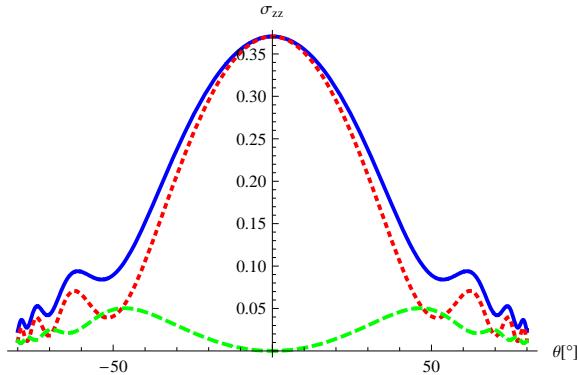}
\caption{{}The angular dependence of conductivity $\protect\sigma _{zz}$,
calculated using Eq. (\protect\ref{sB}) with $\protect\tau _{B}$ given by
Eq. (\protect\ref{tauB}) [solid blue line] and with $\protect\tau _{B}=%
\protect\tau _{0}$ [dotted red line]. The dashed green line gives the
difference between these two curves. The parameters for this plot are $%
k_{F}d=2,~m^{\ast }=2m_{e},~B=5T,~\Gamma _{0}=1K$.}
\label{FigYamB5}
\end{figure}

\begin{figure}[th]
\includegraphics[width=0.4\textwidth]{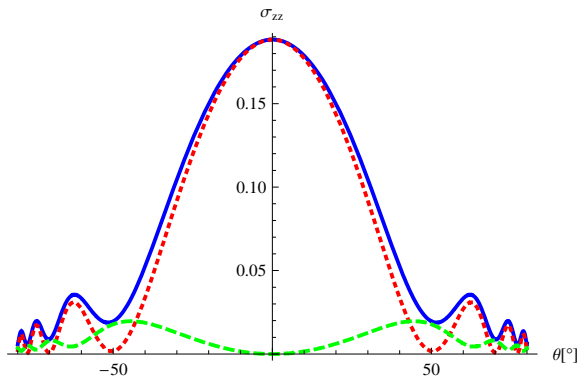}
\caption{{}The angular dependence of conductivity $\protect\sigma _{zz}$,
calculated using Eq. (\protect\ref{sB}) with $\protect\tau _{B}$ given by
Eq. (\protect\ref{tauB}) [solid blue line] and with $\protect\tau _{B}=%
\protect\tau _{0}$ [dotted red line]. The dashed green line gives the
difference between these two curves. The parameters for this plot are the
same as in Fig. \protect\ref{FigYamB5} besides the larger value of magnetic
field $B=20T$. One can see that in the minima of conductivity, i.e. at the
Yamaji angles, the difference is very strong. The predicted value of
conductivity at the Yamaji angles with $\protect\tau _{B}$ given by Eq. (%
\protect\ref{tauB}) is much larger than with $\protect\tau _{B}=\protect\tau %
_{0}$. }
\label{FigYamB20}
\end{figure}

The higher harmonic in Eq. (\ref{sB}) play considerable role in AMRO. To
illustrate this, in Figs. \ref{FigYamB5},\ref{FigYamB20} we plot the angular
dependence of conductivity $\sigma _{zz}\left( \theta \right) $ given by Eq.
(\ref{sB}) with $\tau _{B}=\hbar /2\Gamma _{B}=\tau _{0}\left( \Gamma
_{0}/\Gamma _{B}\right) $ and $\tau _{B}=\tau _{0}$. For simplicity, we take
the axially symmetric case, i.e. the symmetric in plane electron dispersion.
One can see from Figs. \ref{FigYamB5},\ref{FigYamB20} that in the minima of
conductivity, i.e. at the Yamaji angles, the replacement $\tau
_{0}\rightarrow \tau _{B}$ is very important. The predicted value of
conductivity at the Yamaji angles with $\tau _{B}$ given by Eq. (\ref{tauB})
is much larger than with $\tau _{B}=\tau _{0}$. This difference increases
with increasing of magnetic field. The positions of the conductivity minima,
i.e. the Yamaji angles, also slightly shift after the replacement $\tau
_{0}\rightarrow \tau _{B}$ in Eq. (\ref{sB}) [see Figs. \ref{FigYamB5},\ref%
{FigYamB20}]. For the first Yamaji angle at $B=5T$ (see Fig. \ref{FigYamB5})
this shift $\Delta \theta _{Yam}\approx 1.7^{\circ }$. \

\section{Discussion}

Let us formulate the main difference of the present approach to the
calculation of interlayer conductivity in the weakly incoherent regime
compared to the previous methods, developed in Refs. \cite{ShubCondMat}\cite%
{ChampelMineev}\cite{Shub}\cite{Gvozd2004}\cite{ChMineevComment2006} to
calculate the MQO of conductivity. In these papers the impurity potential is
considered as a small perturbation on the background of a free electron gas
with well-defined 3D electron dispersion given by Eq. (\ref{ES3D}). Hence,
the impurity scattering was taken into account only by the imaginary part of
the electron self-energy given by Eq. (\ref{G}). Even less accurately the
impurities are treated in Ref. \cite{MosesMcKenzie1999}, where the constant
electron mean-free time has been used to include the interaction with
impurities. This is correct only in the coherent limit, when the interlayer
transfer integral is much larger than the LL broadening, and the electrons,
moving in a 3D metal, are scattered by impurities. In the weakly incoherent
regime, when $t_{z}<\Gamma $, this is incorrect, because for a 2D electron
system in magnetic field the impurity potential has much stronger effect
than in 3D. Simply, in a 3D electron system the electrons after scattering
by an impurity move away in the interlayer direction and never return to
this impurity. Therefore, this impurity only leads to the single scattering
of this electron into some other state, which is well described by the
constant electron mean-free time $\tau _{0}$, or equivalently, by the
constant imaginary part $\Gamma _{0}$ of the electron self-energy. In 2D
electron system in magnetic field, the electrons after scattering return to
the same impurity after the cyclotron period. Therefore, the impurity has
permanent influence on the electron state, considerably shifting the
electron energy and modifying the electron states. Hence, in the weakly
incoherent regime, when $t_{z}<\Gamma $, the interlayer hopping term (\ref%
{Ht}) in the Hamiltonian (\ref{H}), rather the impurity potential (\ref{Hi}%
), must be considered as a small perturbation. Therefore, to calculate the
interlayer conductivity, we start from the stack of isolated 2D disordered
conducting layers in magnetic field, where the effect of impurity potential
is considered much more accurately, at least in the self-consistent one-site
approximation. Then we substitute the obtained electron Green functions to
the Kubo formula for the tunnelling conductivity between adjacent conducting
layers. The effect of impurities in the final results turned out to be much
stronger than in the previous approaches. Phenomenologically, this
difference can be taken into account by the replacement of the initial level
broadening $\Gamma _{0}$ by the new value given by Eq. (\ref{G}).

One can also obtain Eq. (\ref{GLor}) with the new value of $\Gamma $ given
by Eq. (\ref{GB}) using different arguments. The physical origin of large
DoS broadening in Eq. (\ref{DoSAndo}) is not the finite lifetime $\tau $ of
electron states, with is mathematically described by the imaginary part of
the self-energy Im$\Sigma =\Gamma _{0}=\hbar /2\tau $, as in the 3D limit.
On the 2D layers the LL broadening comes from the energy shift of each
electron state, which is described by the state-dependent real part of the
electron self-energy Re$\Sigma $. The averaging of the electron Green
function in Eq. (\ref{G0}) over the impurity configurations is independent
on each conducting layer, since the impurity distribution is assumed to be
uncorrelated. Then, the coordinate part of the Green function remains of the
form (\ref{G0}) with the bare electron wave functions in numerator [see Eq. (%
\ref{Gg}) and Appendix], but the denominator acquires the real part of
electron self energy, which is distributed with the DoS function $D\left(
E\right) $:
\begin{equation*}
\left\langle G_{R}^{0}({\boldsymbol{r}}_{1},{\boldsymbol{r}}%
_{2},j,\varepsilon )\right\rangle =\int dE\,D\left( E\right) \sum_{n,k_{y}}%
\frac{\Psi _{n,k_{y},j}^{0\ast }(x_{2},y_{2})\Psi
_{n,k_{y},j}^{0}(x_{1},y_{1})}{\varepsilon -E-\hbar \omega _{c}\left(
n+1/2\right) -i\Gamma _{0}}.
\end{equation*}%
The triangular brackets indicate averaging over impurity configurations.
Substituting the approximate Lorentzian DoS distribution, given by Eq. (\ref%
{LorentzDoS}), one can easily perform the integration over $E$ and obtain
\begin{equation}
\left\langle G_{R}^{0}({\boldsymbol{r}}_{1},{\boldsymbol{r}}%
_{2},j,\varepsilon )\right\rangle \approx \sum_{n,k_{y}}\frac{\Psi
_{n,k_{y},j}^{0\ast }(x_{2},y_{2})\Psi _{n,k_{y},j}^{0}(x_{1},y_{1})}{%
\varepsilon -\hbar \omega _{c}\left( n+1/2\right) -i\left( \Gamma
_{0}+\Gamma _{B}\right) }.  \label{G2DN}
\end{equation}%
This Green function differs from Eq. (\ref{G0}) by the increase of the
imaginary self-energy part: $\Gamma _{0}\rightarrow \Gamma _{0}+\Gamma _{B}$
with $\Gamma _{B}$ given by Eq. (\ref{GE}). This is almost equivalent to Eq.
(\ref{GLor}) with $\Gamma _{B}$ given by Eq. (\ref{GB}).

Unfortunately, the proposed analysis considers only the limiting case $%
\Gamma _{B}\gg t_{z}$, when $\Gamma _{B}$ is given by Eqs. (\ref{GE}) or (%
\ref{GB}), but it is not accurate for the intermediate case
$\Gamma _{B}\sim t_{z}$, where the crossover from the coherent to
the weakly incoherent regime takes place. The phenomenological
formula (\ref{GB}) gives only a qualitative dependence $\Gamma
_{B}\left( B_{z}\right) $ in this region. The crossover from the
coherent to the weakly incoherent regime may be driven by the
disorder (impurity concentration) or by magnetic field $B_{z}$.
The latter happens, because with the increase of magnetic field
the effective LL broadening $\Gamma _{B}$ also increases [see Eq.
(\ref{GE})] and at some crossover field $B_{c}\sim
t_{z}^{2}m_{e}^{\ast }c/\Gamma _{0}e\hbar $ it becomes greater
than the interlayer transfer integral $t_{z}$. To calculate the
exact value $B_{c}$ of the crossover field and to describe the
behavior of interlayer conductivity in this region one needs to
calculate the electron Green function in layered metals with
impurities and magnetic field in the crossover region $\Gamma
_{B}\sim t_{z}$. This is an interesting and still open problem.

Above, we have not studied the MQO in the tilted magnetic field. The second
term in the curly brackets in Eq. (\ref{Kubo1}) does not contribute to the
background magnetoresistance, but it affects the MQO. This term amplifies
the MQO and modifies the angular dependence of the MQO amplitude. This
modification is a fine effect which is harder to measure. The angular
dependence of MQO amplitude is also affected by the Zeeman splitting and
possible magnetic ordering.

If the normalized point-like impurity concentration $c_{i}<1$, the $%
N_{LL}-N_{i}$ electron states on each LL left degenerate, and besides the
DoS dome the sharp $\delta \left( E\right) $ term in the DoS survives.\cite%
{Baskin} However, as has been shown in Ref. \cite{Imp}, the
numerous weak impurities and the impurities, situated far from the
conducting layers, are important for the lifting of the LL
degeneracy in all layered materials. For
achievable magnetic field even in the pulsed magnets $B<100T$, $l_{Hz}>25%
\mathring{A}$. Therefore, the typical normalized impurity concentration is
greater than unity, $c_{i}>1$, and one can use the one-maximum DoS
distribution as in Eq. (\ref{DoSAndo}).

Above we have shown that the weakly incoherent regime strongly
differs from the coherent limit. It also differs from the
completely incoherent limit, where the new mechanisms of the
interlayer electron transport, including the electron interlayer
transport via resonance
impurities\cite{Abrikosov1999,Maslov,Incoh2009} and the hopping
conductivity between completely localized states\cite{Gvozd2007},
play important role. One difference of the weakly incoherent
regime from the completely incoherent one is that the angular
magnetoresistance oscillations (AMRO) are not damped, being of the
same amplitude as in the coherent regime. Only higher harmonics in
AMRO increase, making AMRO maxima less pronounced. The second
difference is that the temperature dependence of conductivity in
the weakly incoherent regime is the same, as in the coherent limit
(usually, metallic), while the temperature dependence of the
hopping conductivity\cite{Gvozd2007} is exponential. Therefore,
the weakly incoherent regime of interlayer magnetotransport is a
separate regime, which should be distinguished from the coherent
and completely incoherent limits.

\section{Summary}

We reexamine theoretically the conducting properties of layered
metals in the "weakly incoherent" regime, when the interlayer
transfer integral $t_{z}$ is much less than the Landau level
broadening $\Gamma _{B}$ due to the interaction with impurities.
The magnetic quantum oscillations and the angular dependence of
interlayer conductivity in this regime are calculated. We obtain
that both these effects in the weakly incoherent limit
considerably differ from the coherent regime. This contradicts the
previous theoretical
results.\cite{MosesMcKenzie1999}\cite{ChampelMineev} The
background interlayer conductivity $\sigma_{zz}$ decreases with
the increase of magnetic field $B_{z}$ according to Eq. (\ref{s0})
with $\Gamma _{B}$ approximately given by Eq. (\ref{GB}), while in
the coherent limit it remains constant (see Fig. \ref{FigMQO} for
illustration). The Dingle temperature of MQO also increases with
magnetic field $\propto \Gamma _{B}$. Meantime, in the weakly
incoherent regime the angular oscillations of background
magnetoresistance are not damped as in the completely incoherent
mechanisms of the interlayer
electron transport, considered, e.g., in Ref. \cite{Abrikosov1999},\cite%
{Gvozd2007},\cite{Incoh2009}. On contrary, the damping of higher
harmonics in the angular magnetoresistance oscillations is weaker
than in the coherent regime [see Eqs. (\ref{sB}),(\ref{tauB})].
This leads to the different picture of AMRO (see Figs.
\ref{FigYamB5},\ref{FigYamB20}). Phenomenologically, the
differences between the coherent and weakly incoherent regimes can
be taken into account by the replacement of the electron mean free
time $\tau _{0}$  by the new value $\tau _{B}\approx \tau _{0}/
\left[ \left( 8\omega _{c}\tau _{0}/\pi \right) ^{2}+1\right]
^{1/4}$ in all formulas for MQO and for the angular oscillations
of interlayer conductivity.

\section{Acknowledgement}

The work was supported by GK P1419 of the FCP program "Nauchnye i
Nauchno-Pedagogicheskie Kadry Rossii", by RFBR and by the grant of
the President of Russia MK-2320.2009.2.

\appendix

\section{The in-plane electron Green function in the impurity potential}

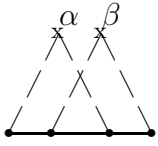
\begin{figure}[tb]
{\small 
}
\par
\begin{center}
{\small
\begin{picture}(60,55)
\multiput(4,5)(7,14){3}{\line(1,2){5}}
\multiput(42,5)(-7,14){3}{\line(-1,2){5}}
\put(23, 46){\large{$\alpha $}}
\put(20, 41){\small{x}}
\put(4, 5){\circle*{3}}
\put(42, 5){\circle*{3}}
\multiput(20,5)(7,14){3}{\line(1,2){5}}
\multiput(58,5)(-7,14){3}{\line(-1,2){5}}
\put(39, 46){\large{$\beta $}}
\put(36, 41){\small{x}}
\put(20, 5){\circle*{3}}
\put(58, 5){\circle*{3}}
\put(4, 5){\line(1,0){53}}
\end{picture}
}
\par
{\small 
}
\end{center}
\caption{ The first diagram for the electron self-energy with the
intersection of the impurity lines. }
\label{FigDiaInt}
\end{figure}

Consider the noninteracting 2D electron gas in the potential of
randomly distributed point-like impurity, as given in Eq.
(\ref{Vi2D}). The peculiarity of the two-dimensional electron gas
in strong magnetic field in the presence of impurities is that the
Born approximation of the scattering amplitude on each impurity is
insufficient to describe the system. Physically, this means that
an electron scatters many times by one impurity, because in
magnetic field the electrons periodically return to the same point
after passing along the cyclotron orbit. In the 3D case the
diagram in Fig. \ref{FigDiaInt} with the intersections of impurity
lines is small by the parameter $n_{i}/n_{e}$, where $n_{i}$ and
$n_{e}$ are the volume impurity and electron concentrations. In 2D
case in magnetic field there is no general proof that the diagrams
with intersections of impurity lines are small. However, the
calculations of the DoS in Refs. \cite{Ando1,Brezin,Burmi} show
that these diagrams only lead to the small tails of the DoS.
Therefore, in the our subsequent analysis we keep only the
diagrams without intersections of impurity lines.
\begin{figure}[tb]
{\small 
}
\par
\begin{center}
{\small
\begin{picture}(120,55)
\put(0, 5){\Large{$\Sigma =$}}
\multiput(30,5)(0,14){3}{\line(0,1){10}}
\put(30, 46){\large{$\alpha $}}
\put(28, 41){x}
\put(30, 5){\circle*{3}}
\put(36, 5){\Large{+}}
\multiput(54,5)(7,14){3}{\line(1,2){5}}
\multiput(92,5)(-7,14){3}{\line(-1,2){5}}
\put(73, 46){\large{$\alpha $}}
\put(70, 41){\small{x}}
\put(54, 5){\circle*{3}}
\put(92, 5){\circle*{3}}
\put(54, 6){\line(1,0){38}}
\put(54, 4){\line(1,0){38}}
\put(96, 5){\Large{+}}
\multiput(114,5)(7,14){3}{\line(1,2){5}}
\multiput(133,5)(0,14){3}{\line(0,1){10}}
\multiput(152,5)(-7,14){3}{\line(-1,2){5}}
\put(133, 46){\large{$\alpha $}}
\put(130, 41){\small{x}}
\put(114, 5){\circle*{3}}
\put(152, 5){\circle*{3}}
\put(133, 5){\circle*{3}}
\put(114, 6){\line(1,0){38}}
\put(114, 4){\line(1,0){38}}
\put(156, 5){\Large{+ ..}}
\end{picture}
}
\par
{\small 
}
\end{center}
\caption{ The set of diagrams for the irreducible self-energy, corresponding
to the self-consistent one-site approximation. The double solid line
symbolizes the exact electron Green's function. }
\label{FigSE}
\end{figure}

Now we proof by the method of mathematical induction, that, if one neglects
the diagrams with the intersection of impurity lines, the electron Green
function, averaged over impurity configurations, has the form of Eq. (\ref%
{Gg}) with
\begin{equation}
G\left( \varepsilon ,n\right) =1/\left[ \varepsilon -\varepsilon _{n}-\Sigma
_{n}\left( \varepsilon \right) \right] .  \label{Gn}
\end{equation}%
The energy $\varepsilon _{n}$ of the $n$-th LL is given by Eq. (\ref{En2D}),
and the electron wave functions $\Psi _{n,k_{y}}^{0}(r)$ by Eq. (\ref{Psi2D}%
). The self energy part $\Sigma _{n}\left( \varepsilon \right) $ for the $n$%
-th LL must be determined self-consistently and is given by the set of
diagrams, shown in Fig. \ref{FigSE}. In the self-consistent one-site
approximation the self-energy part is
\begin{equation*}
\Sigma _{n}\left( \varepsilon \right) =\frac{E-E_{g}\left( 1-c_{i}\right) }{2%
}\mp \frac{\sqrt{\left( E-E_{1}\right) \left( E_{2}-E\right) }}{2}.
\end{equation*}%
The restriction given by Eq. (\ref{Gg}) is nontrivial because $G\left(
\varepsilon ,n\right) $ does not depend on $k_{y}$.

Without impurities, i.e. in the zeroth order of mathematical induction, Eq. (%
\ref{Gg}) holds by definition. Assume, it holds for an arbitrary number $N$
of impurities in the electron Green function $G_{N}({\boldsymbol{r}}_{1},{%
\boldsymbol{r}}_{2},\varepsilon )$. When we add one more impurity center,
the Green function $G_{N+1}({\boldsymbol{r}}_{1},{\boldsymbol{r}}%
_{2},\varepsilon )$ is given by
\begin{equation}
G_{N+1}({\boldsymbol{r}}_{1},{\boldsymbol{r}}_{2},\varepsilon )=\int
dr_{\alpha }G_{N}({\boldsymbol{r}}_{1},{\boldsymbol{r}}_{\alpha
},\varepsilon )G_{N}({\boldsymbol{r}}_{\alpha },{\boldsymbol{r}}%
_{2},\varepsilon )\Sigma \left( \varepsilon ,{\boldsymbol{r}}_{\alpha
}\right) ,  \label{GN1}
\end{equation}%
where $\Sigma \left( \varepsilon \right) $ is given by the set of diagrams
in Fig. \ref{FigSE} with the double-line standing for $G_{N}({\boldsymbol{r}}%
_{\alpha },{\boldsymbol{r}}_{\alpha },\varepsilon )$. Performing the
integration over $k_{y}$ in Eq. (\ref{Gg}), we find
\begin{equation*}
G_{N}({\boldsymbol{r}}_{\alpha },{\boldsymbol{r}}_{\alpha },\varepsilon
)=\sum_{n}\frac{N_{LL}}{\varepsilon -\varepsilon _{n}-\Sigma _{N,n}\left(
\varepsilon \right) }.
\end{equation*}%
Therefore, $\Sigma \left( \varepsilon ,{\boldsymbol{r}}_{\alpha }\right)
=\Sigma \left( \varepsilon \right) $, and substituting Eq. (\ref{Gg}) into
Eq. (\ref{GN1}) and integrating over ${\boldsymbol{r}}_{\alpha }$, we obtain%
\begin{eqnarray}
G_{N+1}({\boldsymbol{r}}_{1},{\boldsymbol{r}}_{2},\varepsilon )
&=&\sum_{n,k_{y}}\frac{\Psi _{n,k_{y}}^{0\ast }(r_{2})\Psi
_{n,k_{y}}^{0}(r_{1})}{\varepsilon -\varepsilon _{n}-\Sigma _{N,n}\left(
\varepsilon \right) }+c_{i}\int dr_{\alpha }\sum_{n,k_{y}}\frac{\Psi
_{n,k_{y}}^{0\ast }(r_{\alpha })\Psi _{n,k_{y}}^{0}(r_{1})}{\varepsilon
-\varepsilon _{n}-\Sigma _{N,n}\left( \varepsilon \right) }\sum_{n^{\prime
},k_{y}^{\prime }}\frac{\Psi _{n^{\prime },k_{y}^{\prime }}^{0\ast
}(r_{2})\Psi _{n^{\prime },k_{y}^{\prime }}^{0}(r_{\alpha })}{\varepsilon
-\varepsilon _{n^{\prime }}-\Sigma _{N,n^{\prime }}\left( \varepsilon
\right) }\Sigma \left( \varepsilon \right) +..  \notag \\
&=&\sum_{n,k_{y}}\frac{\Psi _{n,k_{y}}^{0\ast }(r_{2})\Psi
_{n,k_{y}}^{0}(r_{1})}{\varepsilon -\varepsilon _{n}-\Sigma _{N,n}\left(
\varepsilon \right) }\sum_{j=0}^{\infty }\left( \frac{c_{i}\Sigma \left(
\varepsilon \right) }{\varepsilon -\varepsilon _{n}-\Sigma _{N,n}\left(
\varepsilon \right) }\right) ^{j}=\sum_{n,k_{y}}\frac{\Psi _{n,k_{y}}^{0\ast
}(r_{2})\Psi _{n,k_{y}}^{0}(r_{1})}{\varepsilon -\varepsilon _{n}-\Sigma
_{N+1,n}\left( \varepsilon \right) },  \label{GNp1}
\end{eqnarray}%
where%
\begin{equation*}
\Sigma _{N+1,n}\left( \varepsilon \right) =\Sigma _{N,n}\left( \varepsilon
\right) +c_{i}\Sigma \left( \varepsilon \right) .
\end{equation*}%
Eq. (\ref{GNp1}) has the form (\ref{Gg}), which proves our statement.

\end{document}